\documentclass[useAMS,usenatbib,usegraphicx]{mn2e}

\title[$N$-body Models of Extended Clusters]{$N$-body models of extended star clusters}  
\author[J. R. Hurley \& A. D. Mackey]{Jarrod R. Hurley$^{1}$\thanks{
E-mail: jhurley@swin.edu.au (JRH)}, 
A. Dougal Mackey$^2$ \\
$^{1}$Centre for Astrophysics and Supercomputing, Swinburne University of Technology, P.O. Box 218, VIC 3122, Australia \\
$^{2}$Research School of Astronomy \& Astrophysics, The Australian National University, 
Mount Stromlo Observatory, \\ Cotter Road, Weston Creek, ACT 2611, Australia}

\begin{document}

\date{Accepted 2010 Month xx. Received 2010 Month xx; in original form 2010 May 7} 

\pagerange{\pageref{firstpage}--\pageref{lastpage}} \pubyear{2010}

\maketitle

\label{firstpage}

\begin{abstract}
We use direct $N$-body simulations to investigate the evolution of star clusters with 
large size-scales with the particular goal of understanding the so-called extended clusters 
observed in various Local Group galaxies, including M31 and NGC$\,6822$. 
The $N$-body models incorporate a stellar mass function, stellar evolution and the tidal 
field of a host galaxy. 
We find that extended clusters can arise naturally within a weak tidal field provided that 
the tidal radius is filled at the start of the evolution. 
Differences in the initial tidal filling-factor can produce marked differences in the subsequent 
evolution of clusters and the size-scales that would be observed. 
These differences are more marked than any produced by internal evolution processes 
linked to the properties of cluster binary stars or the action of an intermediate-mass black hole, 
based on models performed in this work and previous work to date. 
Models evolved in a stronger tidal field show that extended clusters cannot form and evolve 
within the inner regions of a galaxy such as M31. 
Instead our results support the suggestion many extended clusters found in large galaxies 
were accreted as members of dwarf galaxies that were subsequently disrupted. 
Our results also enhance the recent suggestion that star clusters evolve to a common sequence 
in terms of their size and mass. 
\end{abstract}


\section{Introduction}
\label{s:intro}

Recent observations have unearthed a new class of extragalactic objects
which have been termed extended star clusters.
These have shown up in the outer regions of a range of galaxy types:
spirals M31 (Huxor et al. 2005; Mackey et al. 2006; Huxor et al. 2008)
and M33 (Stonkut\.{e} et al. 2008; Huxor et al. 2009),
the dwarf irregular NGC$\,6822$ (Hwang et al. 2005; Huxor et al. 2010b)
and the Sculptor Group dwarf elliptical Scl-dE1 (Sc22: Da Costa et al. 2009).
The few extended clusters (ECs) for which high quality colour-magnitude diagrams
(CMDs) are available appear old and metal-poor with no evidence for
multiple populations (Mackey et al. 2006).
Thus they have similar stellar populations to those of typical Milky Way
and M31 globular clusters (GCs). A faint extended cluster in the giant
elliptical galaxy NGC$\,5128$ has also recently been discovered (Mouhcine et al. 2010)
with properties similar to the Palomar GCs found in the outer regions of the
Milky Way (van den Bergh \& Mackey 2004).

The size scales of the ECs,
generally measured as the projected half-light radius, $r_{\rm h,l}$
(where we will use the three-dimensional $r_{\rm h}$ for the related half-mass radius),
are in the range of $10 - 35\,$pc which places them
between regular globular clusters and low-luminosity dwarf spheroidal galaxies
(Huxor et al. 2005; Da Costa et al. 2009).
As such these objects raise questions as to whether they are star clusters that have somehow
evolved to much larger radii than is typical for Milky Way globular clusters or
if they are more closely associated with dwarf spheroidals, perhaps even to the extent
of possessing some kind of dark matter component.
Interestingly there are extended clusters in M31 with the same luminosity,
and at the same distance from the centre of the galaxy, $R_{\rm gc}$, as regular
globular clusters but with a half-light radius a factor of ten or more greater
(Huxor et al. 2010a).
If the extended clusters are indeed a sub-class of globular cluster then we
must ask how clusters that appear very similar in many respects can obtain quite
distinct structural properties.

In the case of two clusters of the same mass $M$ at the same $R_{\rm gc}$ but with
markedly different half-mass radii we must consider the prior $r_{\rm h}$
history of both clusters, possibly traced back to having distinct initial $r_{\rm h}$
values.
In turn this may indicate different modes of cluster formation
(Elmegreen 2008; Da Costa et al. 2009).
Baumgardt \& Kroupa (2007) have shown that variations in star formation efficiency
for clusters as they form from molecular clouds can lead to a range of initial
half-mass radii.
One of the goals of this paper is to investigate the ramifications of this for the
long-term structural evolution of realistic star cluster models.
In particular we would like to know if large differences in the structure of clusters
at a Hubble time is more likely a result of different structures at birth,
different internal evolution histories (e.g. formation of binaries)
or possibly having experienced different external environments in the past.
Of course a combination of these effects is also possible.
We are also interested generally in producing $N$-body models of potential
extended star clusters.

In general terms the evolution of a star cluster is dictated by internal processes,
primarily two-body relaxation,
as well as the influence of the gravitational potential of the host galaxy.
Historically the gross properties, such as the dissolution time,
have been described in terms of $R_{\rm gc}$, $M$ and
%the half-mass radius,
$r_{\rm h}$
(Fall \& Rees 1977; Gnedin \& Ostriker 1997). 
The first two can be used to define a tidal radius, $r_{\rm t}$, while the timescale 
for two-body relaxation can be derived from the latter two quantities. 
Typically the timescale on which 
stars are pushed across the tidal boundary and the cluster subsequently
evaporates owing to two-body relaxation can be 
estimated from $M$ and $r_{\rm h}$ as well 
(see Gnedin, Lee \& Ostriker 1999 for a discussion) 
although as shown by Gieles \& Baumgardt (2008) the tidal field strength 
is the most important factor in determining mass-loss rates. 
For clusters orbiting in the inner regions of a galaxy the lifetime can also
be affected by gravitational shocks from the disk and/or bulge of the galaxy
(Gnedin \& Ostriker 1997).
If we assume that the half-mass radius of a cluster scales with the
tidal radius, as is often done (and noting that $r_{\rm t}$ decreases with time),
then two clusters of the same mass at the same $R_{\rm gc}$ will
have the same $r_{\rm h}$.
In this picture any differences in $r_{\rm h}$
(assuming the clusters are the same age)
requires that the masses in fact be different or that the clusters have
followed different evolutionary paths.
It may be possible that both clusters did not always reside at $R_{\rm gc}$,
having different galactic orbits, with the extreme scenario being that
one cluster was accreted during a galaxy merger
(Mackey \& Gilmore 2004; Da Costa et al. 2009).
Another possibility is that
internal factors such as the fraction of binaries (Gao et al. 1991) or the
formation of an intermediate-mass black hole (Kormendy \& Richstone 1995)
played a role.
Prior $N$-body models that have documented the effect of such factors on the
evolution of a star cluster include Giersz \& Heggie (1997),
de la Fuente Marcos (1997),
Portegies Zwart et al. (2001),
Baumgardt, Makino \& Hut (2005)
and Mackey et al. (2008).

We will be looking at the evolution of $N$-body models that start with different
sizes and models that evolve in tidal fields of different strength.
For the latter we have in mind M31 and NGC$\,6822$ as examples of
galaxies imposing a strong and weak galactic tidal field, respectively.
We will be interested in the maximum half-mass (and half-light) radii that can be obtained
at the same $R_{\rm gc}$ in these two galaxy types.
In particular, we choose $R_{\rm gc} = 10\,$kpc.
This is motivated by the location of the ECs in NGC$\,6822$ which extend out to $10.8\,$kpc
in projection from the galaxy centre (Hwang et al. 2005; Huxor et al 2010b)
as well as the observation of ECs and GCs of similar luminosity/mass at
about this position in M31. This radius could be considered an extreme choice for
M31 which has both ECs and GCs out to at least $100\,$kpc in projection (Martin et al. 2006,
Huxor et al. 2008, Huxor et al. 2010c), with only the innermost presently-known ECs
residing at $\sim 10\,$kpc; however, selecting this radius for our M31 models will give
an indication of the maximum expected influence of a strong tidal field on EC evolution.
Note also that our models should encompass the situation
for additional ECs such as the two known in M33, which reside in a galaxy intermediate
in mass between M31 and NGC$\,6822$ but again at comparable radii:
$12.4$ and $28.6\,$kpc in projection, respectively (Huxor et al. 2009).
In this sense we expect our results to be quite widely applicable.

The remainder of this paper is structured as follows.
In Section~\ref{s:models} we describe the setup of the $N$-body models,
including the $N$-body software and how the tidal field is modelled.
This is followed by the results in Section~\ref{s:results} and a discussion of the
implications for observations of ECs in Section~\ref{s:discus}.

\section{Models}
\label{s:models}

The model clusters in this work are evolved using the 
{\tt NBODY6} direct $N$-body code (Aarseth 2003). 
Gravitational forces are computed using a fourth-order Hermite integration scheme, 
without softening, on either standard CPU or graphics processing unit (GPU) 
hardware. 
{\tt NBODY6} is the sister code to {\tt NBODY4} which instead interfaced 
with GRAPE-6 hardware (Makino 2002) for rapid force calculation. 
The codes include algorithms for stellar and binary evolution as described in
Hurley (2008a, 2008b) and deal directly with the effects of close dynamical encounters:
perturbations to binary orbits, collisions and mergers, formation of three- and
four-body subsystems, exchange interactions, tidal capture and binary disruption. 

We focus on a set of models starting with $N = 100\,000$ particles (stars or binaries). 
Each model includes a mass function, stellar evolution and the tidal field of a 
parent galaxy. 
Single star masses are taken from the initial mass function (IMF) 
of Kroupa, Tout \& Gilmore (1993) between the limits of $0.1 - 30 \, M_\odot$. 
When including binary stars the masses of two randomly chosen single stars 
are combined and then reassigned using a mass-ratio from a uniform distribution 
in order to give the primary and secondary masses. 
Initial periods are drawn from the log-normal distribution determined by 
Duquennoy \& Mayor (1991) from observations of local solar-type stars and 
the eccentricities are assumed to follow a thermal distribution (Heggie 1975). 
Metallicity is set at $Z = 0.001$ for the stars. 
The initial positions and velocities are assigned according to a either a Plummer 
density profile (Plummer 1911; Aarseth, H\'{e}non \& Wielen 1974) or a 
King model (King 1966: $W_0 = 7$) with the stars in virial equilibrium. 

The influence of the tidal field of the parent galaxy is modelled by assuming that 
the model cluster follows a circular orbit at distance $R_{\rm gc}$ from a point-mass 
galaxy of mass $M_{\rm g}$. 
At this stage we are not interested in the complicating factors presented 
by more realistic three-dimensional galaxy potentials. 
The implications of this decision will be discussed in Section~\ref{s:discus}. 
The relevant equations of motion are developed by linearizing the galactic potential 
in a rotating reference frame centered on the cluster centre-of-mass with the 
$x$-axis directed away from the galactic centre and the $y$-axis in the 
direction of motion (see Giersz \& Heggie 1997; Vesperini \& Heggie 1997). 
A tidal radius (also called the Jacobi radius: Gieles \& Baumgardt 2008), 
\begin{equation}\label{e:rtide} 
r_{\rm t} = \left( M / 3 M_{\rm g} \right)^{1/3} \, R_{\rm gc} \, , 
\end{equation} 
can then be defined corresponding to the saddle point on the $x$-axis 
of the effective cluster potential. 
Within an $N$-body simulation the user has the freedom to set a length-scale $R_{\rm sc}$ 
with one possible choice being that the outermost stars of the initial density profile 
are scaled to sit at $r_{\rm t}$, in which case the cluster is said to be Roche-lobe filling 
(e.g. Tanikawa \& Fukushige 2005). 
The Plummer profile formally extends to infinite radius so in practice a cut-off at a radius 
of $\sim 10 \, r_{\rm h}$ is applied to avoid rare cases of large distance. 
For our Plummer models described below this leads to $r_{\rm max} \simeq 8 \, r_{\rm h}$,  
where $r_{\rm max}$ is the position of the outermost star, and we define 
$r_{\rm max} / r_{\rm t}$ as the tidal-radius filling factor. 
The related ratio $r_{\rm h} / r_{\rm t}$ is an important quantity to describe 
the structure of star cluster models. 

In this work we model parent galaxies with two distinct masses: 
$M_{\rm g} = 9 \times 10^{9} \, M_\odot$ to represent a dwarf galaxy such as NGC$\,6822$ 
and $M_{\rm g} = 9 \times 10^{10} \, M_\odot$ to model a more substantial galaxy 
such as M31. 
As discussed above all clusters are set to orbit at $R_{\rm gc} = 10\,$kpc in these model galaxies. 

Six distinct models are performed. 
These are listed in Table~\ref{t:table1}. 
Models N1, N2 and N3 are all evolved in the NGC$\,6822$-like tidal field. 
They are identical in all respects except for using $R_{\rm sc} = 7$, 14 and 21 so that 
they have initial tidal-radii filling factors of 0.33, 0.66 and 1.00, respectively 
(the corresponding $r_{\rm h} / r_{\rm t}$ ratios are given in Table~\ref{t:table1}). 
Model N2b is the same as N2 except that it starts with $95\,000$ single stars and 
$5\,000$ binaries rather than $100\,000$ single stars. 
We then have Model M1 which is evolved in the stronger M31-like tidal field 
for comparison. 
All of these models start with a Plummer density profile whereas Model M2 starts 
with a King profile and is also evolved in the M31-like tidal field. 
Both M1 and M2 start with tidal-radii filling factors of 1.00 ($r_{\rm max}$ matches the tidal radius). 
The initial mass of each model is $M \simeq 58\,000 \, M_\odot$ which gives an initial 
$r_{\rm t}$ of $129\,$pc for models N1, N2, N2b and N3 compared to $60\,$pc 
for models M1 and M2. 
Each model is evolved to an age of $20\,$Gyr or until  5\% of the stars 
remain, whichever occurs first 
-- the latter only happens for M2 at an age of $17.4\,$Gyr. 
The simulations are performed using Tesla S1070 GPUs at Swinburne University. 

We will also lean on the results of some previous $N$-body simulations when 
evaluating our results. 
These include models K100-00a and K100-00b of Hurley (2007) which both 
featured $100\,000$ single stars evolved within a standard Galactic tidal field 
(Giersz \& Heggie 1997): 
an orbital speed of $220 \, {\rm km} \, {\rm s}^{-1}$ at $R_{\rm gc} = 8.5\,$kpc 
(with corresponding $M_{\rm g} = 9 \times 10^{10} \, M_\odot$). 
The two models were setup in the same way and the difference of note was the 
formation of a long-lived binary composed of two stellar-mass black holes (BHs) 
in K100-00b which altered the central 
structure of the cluster compared to K100-00a. 
Along the same lines we will use the results of Mackey et al. (2008) who looked at the 
effect of a population of BH-BH binaries on cluster evolution, 
as well as models including intermediate-mass black-holes (e.g. Gill et al. 2008). 
Also mentioned will be models of $30\,000$ single stars from Hurley et al. (2004), 
mainly for illustrative purposes. 

It is important to emphasize that unless otherwise specified the radii quoted will 
be based on three-dimensional data. 
The core radius, $r_{\rm c}$, comes from a density-weighted calculation (Casertano \& Hut 1985) 
that is traditionally used in $N$-body models and is not comparable to the quantity 
derived by observational methods. 
This has been discussed in the past (e.g. Wilkinson et al. 2003; Hurley 2007). 
For our purposes this is fine as we use $r_{\rm c}$ as an indicator of the cluster dynamical 
state, in particular to determine if core-collapse has been reached, rather than to 
compare to observed results for actual clusters. 
We also use $r_{\rm h}$ as the 
three-dimensional half-mass radius. 
Here we are mainly interested in the relative values between models. 
However, we will also provide the half-light radius, $r_{\rm h,l}$, calculated from a 
two-dimensional projection of the data, to give a reference point for comparing 
the size of the model clusters to real clusters. 

\begin{table*}
\begin{minipage}{126mm}
\caption{
Key information for the initial models of the new simulations performed in this work 
and also the models K100-00a and K100-00b from Hurley (2007). 
All models started with $N = 100\,000$ stars and $M \simeq 58\,000 \, M_\odot$. 
Column~1 gives the label assigned to each model where N represents the 
NGC$\,6822$-like tidal field and M the M31-like tidal field. 
In Columns~2 and 3 we show the density profile and the primordial binary fraction, 
respectively. 
Columns~4-8 give the half-mass radius, tidal radius, half-mass to tidal radius ratio, 
tidal radius filling factor and half-mass relaxation 
timescale at $T = 0\,$Gyr. 
All lengths are in pc units and relaxation times are in Myr. 
\label{t:table1}
}
\begin{tabular}{llrrrrcr}
\hline  
Label & profile & $f_{\rm b}$ & $r_{\rm h}$ & $r_{\rm t}$ &  $r_{\rm h} / r_{\rm t}$ & $r_{\rm max} / r_{\rm t}$ & $T_{\rm rh}$  \\
\hline
N1 & Plum &  0.00 & 5.4 & 129.0 & 0.042 & 0.33 & 980  \\
N2 & Plum &  0.00 & 10.9 & 129.0 & 0.084 & 0.66 & 2770  \\
N2b & Plum &  0.05 & 10.9 & 129.0 & 0.084 & 0.66 & 2697  \\
N3 & Plum &  0.00 & 16.3 & 129.0 & 0.126 & 1.00 & 5090  \\
M1 & Plum &  0.00 & 7.8 & 60.0 & 0.130 & 1.00 & 1670  \\
M2 & King &  0.00 & 8.7 & 60.0 & 0.145 & 1.00 & 1990  \\
K100-00a & Plum &  0.00 & 6.6 & 51.0 & 0.129 & 1.00 & 1440  \\
K100-00b & Plum &  0.00 & 6.6 & 51.0 & 0.129 & 1.00 & 1430  \\
\hline
\end{tabular}
\end{minipage}
\end{table*}

\begin{table*}
\begin{minipage}{126mm}
\caption{
As for Table~\ref{t:table1} but now showing properties of the models at $T = 12\,$Gyr.  
Column~1 gives the label assigned to each model. 
Column~2 gives the half-mass radius followed by an estimate of the error in this value. 
This is repeated in columns~4 and 5 for the 2-dimensional projected half-light radius. 
The values in columns 2 and 4 are computed from the average of five snapshots 
taken over a $300\,$Myr period centred on $12\,$Gyr. 
Errors in columns 3 and 5 are based on the difference between the average and 
the greatest outlier. 
Columns 6-8 give the tidal radius, 
the mass that remains bound within the tidal radius 
and the half-mass relaxation timescale. 
All lengths are in pc units, masses in $M_\odot$ and relaxation times are in Myr. 
\label{t:table2}
}
\begin{tabular}{lrrrrrrr}
\hline 
Label & $r_{\rm h}$ & error & $r_{\rm h,l}$ & error & $r_{\rm t}$ & $M$ & $T_{\rm rh}$  \\
\hline
N1 & 10.5 & $\pm 0.1$ & 5.6 & $\pm 1.1$ & 106.0 & $32\,470$ & 3240 \\
N2 & 17.3 & $\pm 0.1$ & 10.5 & $\pm 1.0$ & 103.0 & $29\,940$ & 6680 \\
N2b & 17.3 & $\pm 0.1$ & 11.1 & $\pm 1.8$ & 103.0 & $29\,960$ & 6400 \\
N3 &  23.0 & $\pm 0.1$ & 15.4 & $\pm 1.0$ & 97.0 & $25\,410$ & 9440 \\
M1 & 8.1 & $\pm 0.2$ & 5.6 & $\pm 1.2$ & 38.4 & $15\,430$ & 1410 \\
M2 & 5.6 & $\pm 0.1$ & 3.9 & $\pm 1.4$ & 32.5 & $9\,420$ & 620 \\
K100-00a & 5.9 & $\pm 0.1$ & 4.0 & $\pm 1.7$ & 35.4 & $15\,780$ & 890 \\
K100-00b & 6.5 & $\pm 0.1$ & 3.2 & $\pm 0.7$ & 35.2 & $15\,470$ & 1020 \\
\hline
\end{tabular}
\end{minipage}
\end{table*}

\begin{figure}
\includegraphics[width=84mm]{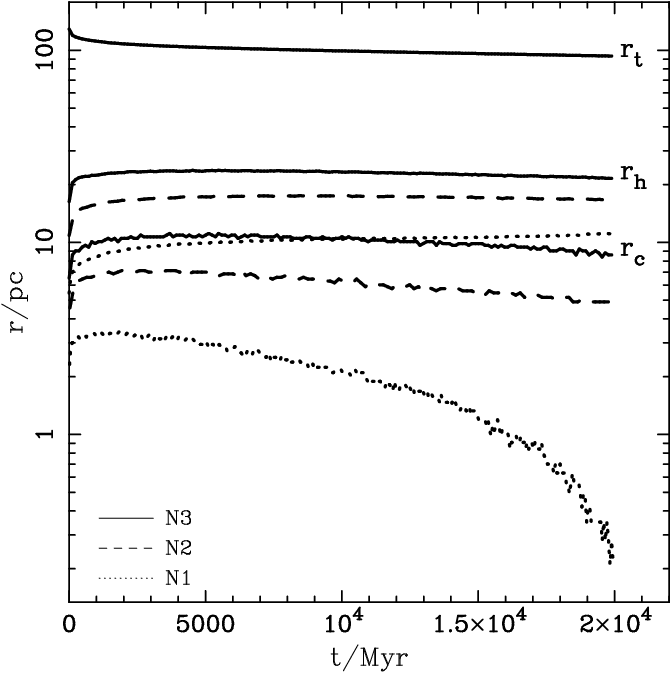}
\caption{
Evolution of the half-mass radius, $r_{\rm h}$, and the core-radius, $r_{\rm c}$,  
for the NGC$\,6822$ models N1 (dotted lines), N2 (dashed lines) and N3 (solid lines) 
noting that $r_{\rm c}$ is always less than $r_{\rm h}$. 
The tidal radius of model N3 is also shown. 
\label{f:fig1}}
\end{figure}

\section{Results}
\label{s:results}

The different initial filling factors of models N1, N2 and N3 lead to 
different initial half-mass radii (see Table~\ref{t:table1}) and allow us to 
investigate the effect this has on the long-term evolution of star clusters 
residing within a weak tidal field. 
It is well established that the evolution of a star cluster is intricately linked 
to the two-body relaxation timescale which is typically characterized by the 
value at the half-mass radius, 
\begin{equation} 
t_{\rm rh} \propto \frac{N}{\log 0.4 N} \left( \frac{r_{\rm h}^3}{M} \right)^{1/2} 
\end{equation} 
(Spitzer 1987; Binney \& Tremaine 1987). 
For a cluster to reach the dynamically-evolved state of core-collapse it has 
been shown that at least ten half-mass relaxation times must have elapsed 
(e.g. Baumgardt, Hut \& Heggie 2002; Hurley 2007). 
The half-mass relaxation timescales corresponding to the initial states of the 
clusters are given in Table~\ref{t:table1} for our models. 
In Table~\ref{t:table2} the timescales for the clusters at an age of $12\,$Gyr are given.  
We can see that of models N1, N2 and N3 it is only model N1, the initially 
most compact cluster, that is likely to reach core-collapse within 
$20\,$Gyr of evolution. 
This is borne out in Figure~\ref{f:fig1} which shows the evolution of  
$r_{\rm c}$ and $r_{\rm h}$ for the three models. 
The tidal radius of model N3 is also shown for reference. 
Indeed, model N1 is close to core-collapse when the simulation is stopped 
at $20\,$Gyr. 
By comparison, the core radii of models N2 and N3 are showing no signs of 
decreasing to the small values associated with core-collapse (at $12\,$Gyr 
$r_{\rm c}$ for N3 is the same size as $r_{\rm h}$ for N1). 
Based on the half-mass relaxation timescales such slow evolution for models 
N2 and N3 is expected as at $20\,$Gyr they have evolved for less than four 
and three half-mass relaxation times, respectively. 

It can be seen in Figure~\ref{f:fig1} that for all three models there is an initial 
rapid phase of increasing $r_{\rm h}$. 
This corresponds to the phase of violent relaxation associated with stellar evolution 
mass-loss from massive stars in which there is an overall expansion of the cluster. 
Afterwards the evolution of $r_{\rm h}$ proceeds slowly and for models N2 and N3 
$r_{\rm h}$ is effectively constant over Gyr timescales. 
In model N1 we do see a continual increase in $r_{\rm h}$ as the outer regions of 
the cluster expand in response to the shrinking core. 
At all times $r_{\rm h}$ for N3 remains larger than for N2 which in turn remains 
larger than for N1: the initial hierarchy of cluster size is maintained and the initially 
more extended cluster remains so. 
It is true that the spread in $r_{\rm h}$ decreases with age 
-- $r_{\rm h}$ for N3 is initially a factor of three greater than for N1 compared 
to a factor of two greater at $20\,$Gyr 
-- but even so the initial cluster size scale has an important bearing on the subsequent 
evolution. 

\begin{figure*}
\includegraphics[width=168mm]{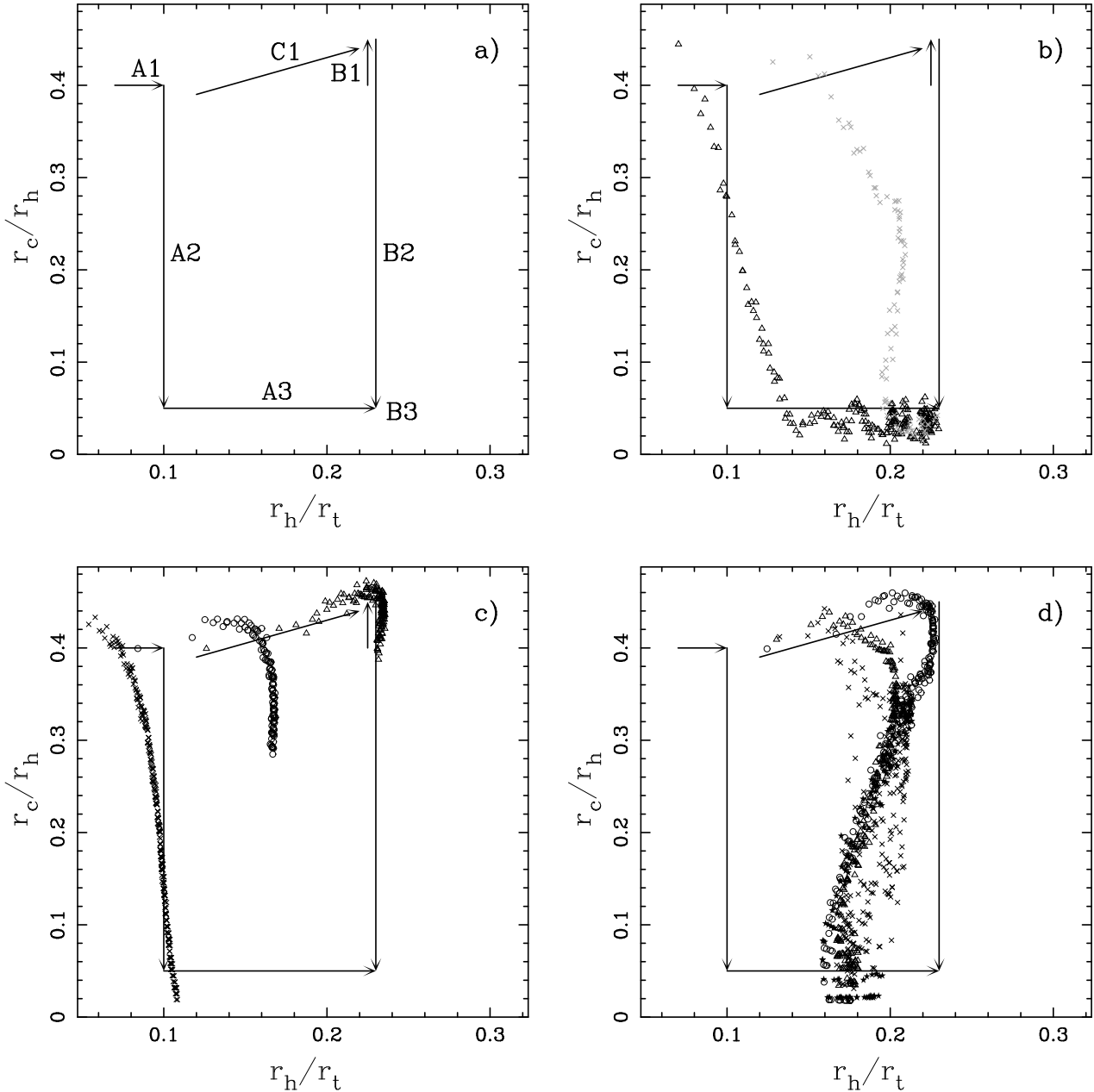}
\caption{
Evolution of star clusters in the $r_{\rm h} / r_{\rm t}$ versus $r_{\rm c} / r_{\rm h}$ plane. 
Panel a) is for illustrative purposes (see text for an explanation of the labels) 
while panel b) shows Models 3 (open triangles) and 7 (grey crosses) of Hurley et al. (2004). 
In panel c) we show models N1 (crosses), N2 (open circles) and N3 (open triangles) of this work. 
Finally, panel d) shows models M1 (open circles) and M2 (solid stars) of this work as well 
as Models K100-00a (open triangles) and K100-00b (crosses) of Hurley (2007). 
\label{f:fig2}}
\end{figure*}

As described by Heggie \& Hut (2003) an interesting way to understand the evolution of a 
star cluster is through ratios of the key radii ($r_{\rm c}$, $r_{\rm h}$ and $r_{\rm t}$). 
In Figure~\ref{f:fig2}a we reproduce Fig.~33.2 of Heggie \& Hut (2003) and use it to 
give a basic overview of the expected behaviour (see Ch.~33 of Heggie \& Hut 2003 
for more background and detail). 
The evolution paths are distinct for clusters initially contained well within the 
tidal radius (path A in Fig.~\ref{f:fig2}a) or filling the tidal radius (path B). 
For the former case all regions of the cluster expand during the initial period of 
violent relaxation so that $r_{\rm c} / r_{\rm h} \sim \,$constant while 
$r_{\rm h} / r_{\rm t}$ increases (as $r_{\rm t}$ changes only minimally). 
Thus the cluster evolves along path A1. 
For the tidally-filling cluster, where in the extreme we assume that $r_{\rm h}$ 
follows the evolution of $r_{\rm t}$, we would instead expect an increase 
in $r_{\rm c} / r_{\rm h}$ while $r_{\rm h} / r_{\rm t}$ remains constant (path B1). 
Following violent relaxation the cluster then enters a long-lived period of dynamical evolution driven by 
two-body relaxation where mass-segregation develops and the cluster evolves 
towards core-collapse. 
During this time $r_{\rm c}$ decreases markedly, $r_{\rm h}$ changes only slightly by comparison 
and $r_{\rm t}$ decreases slowly. 
This gives paths A2 and B2 as both regimes head towards low $r_{\rm c} / r_{\rm h}$. 
Post-collapse evolution is characterized by a central energy source (e.g. hard binaries) 
heating the core which will oscillate in size but with a time-average that varies little. 
For an isolated cluster we would expect $r_{\rm h}$ to increase during this phase 
so any cluster that is not yet tidally-filling will move along path A3 to join B3. 
The tidally-filling clusters are expected to have little evolution in the 
$r_{\rm c} / r_{\rm h}$--$r_{\rm h} / r_{\rm t}$ plane after core-collapse. 
We can readily verify this by looking at the evolution of some actual models. 
First we start with two models of $N = 30\,000$ stars from Hurley et al. (2004) 
where one model started well within $r_{\rm t}$ and the other started with 
$r_{\rm max} = r_{\rm t}$. 
These are shown in Figure~\ref{f:fig2}b and we clearly distinguish path A and path B-like 
evolution before both clusters end at a similar point. 
In reality we would expect most clusters to evolve between the two extremes 
of paths A and B as they head towards B3, particularly during violent relaxation 
where clusters will expand and can move along path C1 to join B2. 
We see this to some extent when we look at the evolution of models N1, N2 and N3 
in Figure~\ref{f:fig2}c. 
Model N3 in particular first evolves across the phase space before starting 
down path B2. 
There is also a clear distinction between the evolution of the three models, 
residing at well separated $r_{\rm h} / r_{\rm t}$ values as they start to 
move down in $r_{\rm c} / r_{\rm h}$. 
Of course, owing to the large half-mass relaxation timescales, the models 
do not get the opportunity to evolve completely through the parameter space. 

\begin{figure}
\includegraphics[width=84mm]{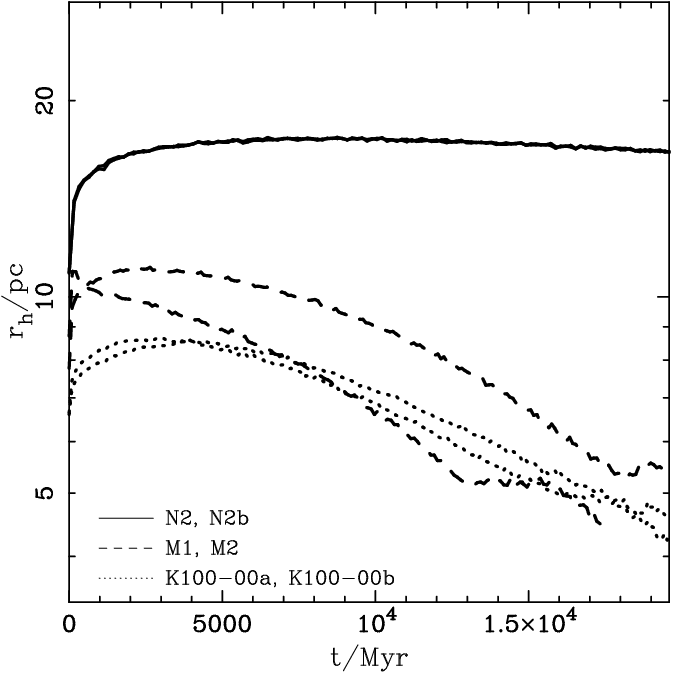}
\caption{
Comparison of the half-mass radius evolution for models N2 and N2b (solid lines), 
M1 and M2 (dashed lines), and K100-00a and K100-00b (dotted lines) of Hurley (2007). 
\label{f:fig3}}
\end{figure}

We have now seen that it is possible to reach large $r_{\rm h}$ values for clusters 
evolving in a weak NGC$\,6822$-like tidal field and that quite distinct $r_{\rm h}$ values  
can be obtained by clusters with different initial sizes. 
But can distinct $r_{\rm h}$ values be obtained by other means? 
The first possibility we explore is the inclusion of a primordial binary population (Model N2b). 
Figure~\ref{f:fig3} compares the $r_{\rm h}$ evolution of models N2 and N2b which are 
identical in setup except for a 5 per cent primordial binary population in the latter. 
We see that the $r_{\rm h}$ evolution is indistinguishable. 
This is also true for the evolution of bound cluster mass (see Table~\ref{t:table2}) 
and other general quantities. 
We note that as an accuracy check models N2 and N2b were both performed twice with different initial 
random number seeds and the variation between different realisations of the same model 
was less than the difference between the two model types. 
For these extended clusters it is not surprising that the addition of binaries 
makes little difference to the evolution. 
The long relaxation times for Models N2 and N2b 
(see Tables~\ref{t:table1} and \ref{t:table2}) 
mean that the collisional extraction of binding energy from the binary 
orbits will not be efficient at heating the cluster. 

The second possibility for internal evolution creating $r_{\rm h}$ differences between models 
relates to the formation of tight BH-BH binaries which then act as a central energy source to heat 
the cluster. 
Hurley (2007) showed that the formation of one such long-lived BH-BH binary could 
double the $r_{\rm c} / r_{\rm h}$ ratio compared to a similar model which did not form 
a BH-BH binary. 
However, we compare the $r_{\rm h}$ evolution of these models (K100-00a and K100-00b from 
Hurley 2007) in Figure~\ref{f:fig3} (and in Tables~\ref{t:table2}) and we see no clear distinction. 
Mackey et al. (2008) took this further and contrasted the evolution of clusters with no BH-BH 
binaries to that of clusters which retained a large number of post-supernovae BHs ($\sim 200$) that 
subsequently sank to the cluster centre and formed BH-BH binaries 
(as many as five such binaries present at any one time). 
The focus was on star clusters in the Large Magellanic Cloud and as such a tidal field the same as for 
our NGC$\,6822$ case was used 
(but with $R_{\rm gc} = 6\,$kpc rather than $10\,$kpc). 
They found that the inclusion of the BH-BH binaries could increase $r_{\rm h}$ by as 
much as a factor of two by the time that the model with no BH-BH binaries had reached 
core-collapse (compared to a corresponding factor of 20 increase in $r_{\rm c}$). 
The $r_{\rm h}$ behaviour first started to diverge after $1-2\,$Gyr of evolution, 
corresponding to roughly one half-mass relaxation time, with the expansion 
driven on the shorter relaxation timescale of the centralised BH population. 
However, we note that the most extended of these models were in the advanced stages  
of dissolution at a Hubble time.  
Another possibility is one that has gathered much attention of late, namely the question 
of whether or not some star clusters harbour intermediate-mass black holes (IMBHs). 
Gill et al. (2008) compare the $r_{\rm h}$ evolution of models with and without an IMBH 
and find no significant difference until well after core-collapse and even at very 
late times the difference is still less than a factor of two. 
Baumgardt, Makino \& Hut (2005) looked at the effect of increasing IMBH mass and found an 
increase in $r_{\rm h}$ of 15 per cent at most. 
It should be noted that the maximum black hole mass included in the models to date 
is $1\,000 \, M_\odot$ and that the heating produced by significantly more massive IMBHs, 
if indeed they exist, is yet to be documented. 

\begin{figure}
\includegraphics[width=84mm]{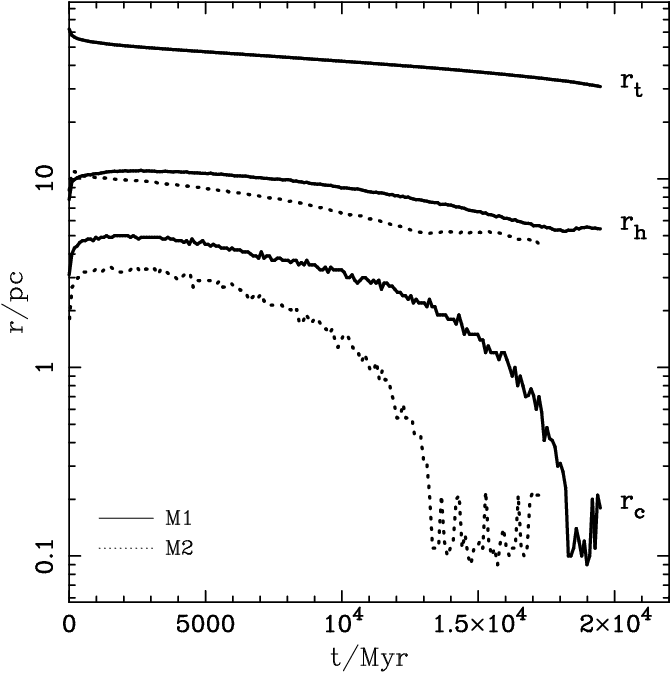}
\caption{
Evolution of the half-mass radius, $r_{\rm h}$, and the core-radius, $r_{\rm c}$,  
for the M31 models M1 (solid lines) and M2 (dashed lines). 
The tidal radius of model M1 is also shown. 
\label{f:fig4}}
\end{figure}

\begin{figure}
\includegraphics[width=84mm]{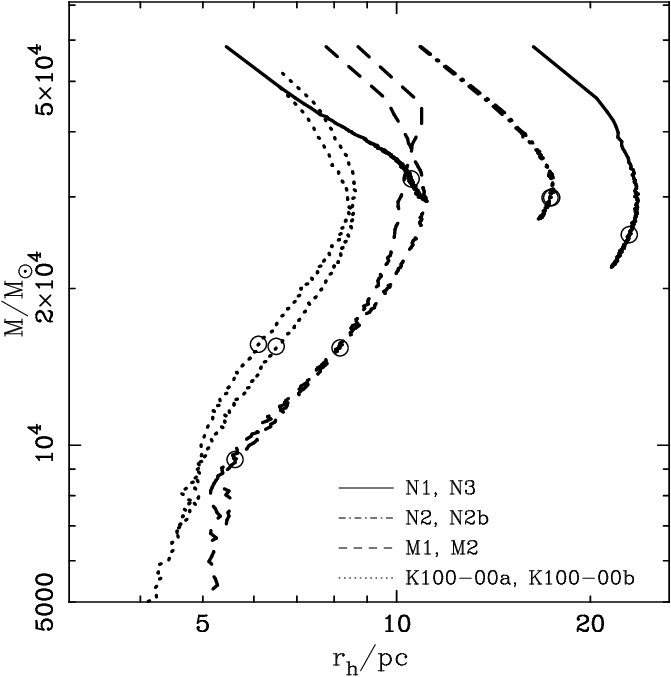}
\caption{
Cluster mass versus half-mass radius for the models in this work 
as well as models K100-00a and K100-00b of Hurley (2007). 
For each evolution track the point at $12\,$Gyr is denoted by an open circle. 
\label{f:fig5}}
\end{figure}

We next look at the evolution of our models M1 and M2 which were evolved in the 
stronger M31-like tidal field. 
These models started with different initial density profiles so provide an opportunity 
to look at how the choice of a Plummer or King profile affects the $r_{\rm h}$ evolution. 
This is shown in Figure~\ref{f:fig3} (also in Figure~\ref{f:fig4}) 
and we see that at various stages in the evolution the difference can be up to 50 per cent. 
The radius evolution of models M1 and M2 is studied in more detail in Figure~\ref{f:fig4}. 
Comparing this to Figure~\ref{f:fig1} we clearly see that the stronger tidal field drives 
more rapid evolution for the M31 models relative to their NGC$\,6822$ counterparts. 
Indeed, both M1 and M2 reach core-collapse prior to $20\,$Gyr. 
The model with the King density profile evolves more rapidly. 
This is primarily owing to a greater central density of stars in the initial model which 
led to a greater rate of dynamical interactions, more mass lost across the tidal boundary 
in the early stages, and consequently a reduced relaxation timescale. 
However, in the $M$-$r_{\rm h}$ plane (Figure~\ref{f:fig5}) the evolution of models 
M1 and M2 is very similar. 
Interestingly both M1 and M2 appear to keep $r_{\rm h}$ approximately constant 
after core-collapse, behaviour that was noted by K\"{u}pper, Kroupa \& Baumgardt (2008) 
in their models. 
We also see that models M1 and M2, as well as models K100-00a and K100-00b 
of Hurley (2007) all evolve towards a similar value of $r_{\rm h} \simeq 5\,$pc even 
though they start with distinct initial conditions or follow different evolution paths in 
other respects. 
The evolution in the $r_{\rm c} / r_{\rm h}$--$r_{\rm h} / r_{\rm t}$ plane is also shown 
for these models in Figure~\ref{f:fig2}d where they all lead to a final 
$r_{\rm h} / r_{\rm t} \simeq 0.18$ and appear to evolve intermediate to the 
N2 and N3 cases (although for N3 it is not dynamically evolved enough to determine 
if the path is distinct from M1 or not). 

\begin{figure}
\includegraphics[width=84mm]{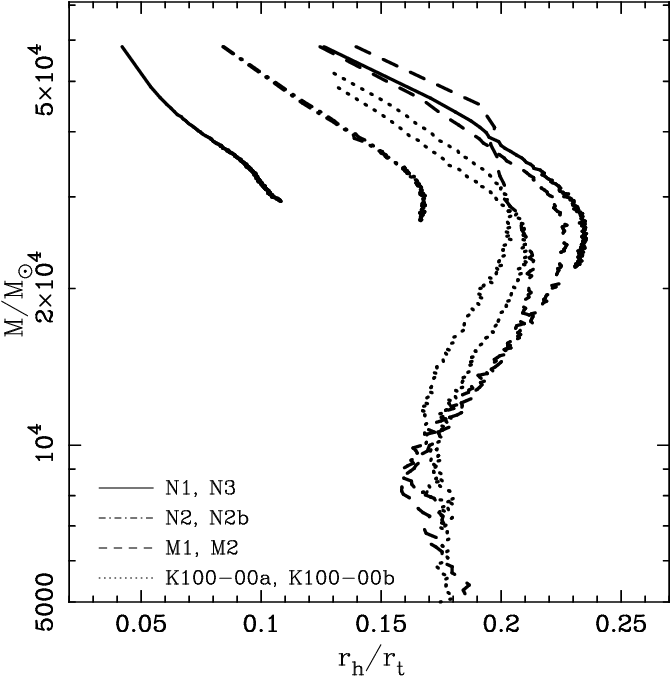}
\caption{
Cluster mass versus the ratio of the half-mass to tidal radius for the models in this work 
as well as models K100-00a and K100-00b of Hurley (2007). 
\label{f:fig6}}
\end{figure}

Figure~\ref{f:fig5} looks at the evolution tracks of our model clusters in 
the $M$--$r_{\rm h}$ plane. 
We see that for any particular mass the initially extended clusters in the weaker 
tidal field have the greater half-mass radii. 
Therefore, clusters at the same $R_{\rm gc}$ in M31 and NGC$\,6822$ will follow 
distinct paths in the $M$--$r_{\rm h}$ plane, assuming they were all close to filling 
their tidal radii initially. 
This is to be expected. 
However, the situation changes if we look at the cluster mass versus $r_{\rm h} / r_{\rm t}$ 
ratio as in Figure~\ref{f:fig6}. 
Now the tidally-filling clusters in M31 and NGC$\,6822$ follow very similar evolution paths. 
K\"{u}pper, Kroupa \& Baumgardt (2008) showed that clusters evolve towards a common 
post-collapse sequence in this plane. 
Our models would appear to back this finding (a point we will return to in Section~\ref{s:discus}) 
and the sequence in our case corresponds to a constant or slightly increasing $r_{\rm h} / r_{\rm t}$ 
ratio. 
For models N1 and N2 it is too early to tell if they too will join the sequence when they 
eventually reach the post-collapse regime but indications are that this is likely. 

\begin{figure}
\includegraphics[width=84mm]{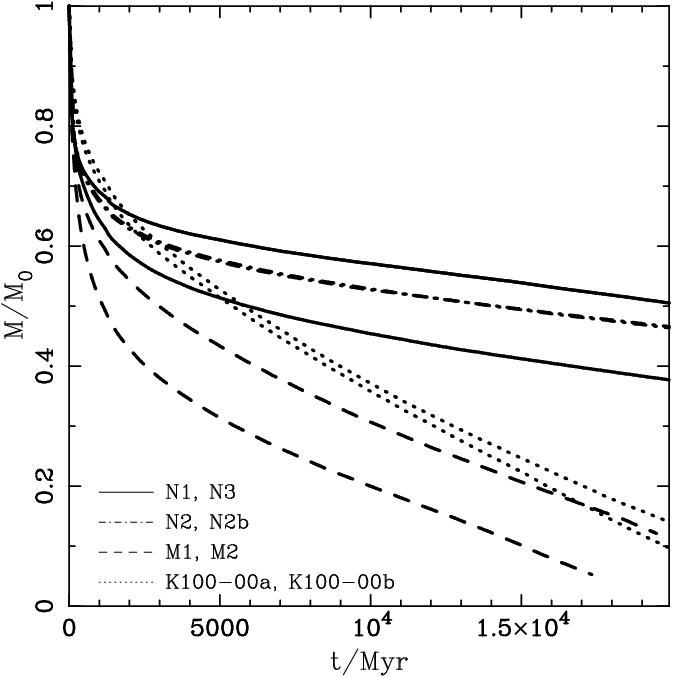}
\caption{
Cluster mass (scaled by the initial mass) as a function of age for the models in this work 
as well as models K100-00a and K100-00b of Hurley (2007). 
\label{f:fig7}}
\end{figure}

\begin{figure}
\includegraphics[width=84mm]{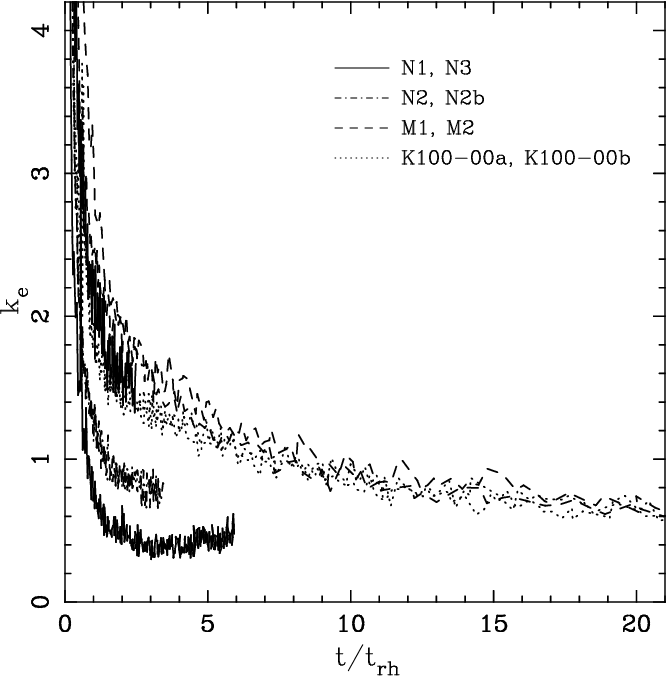}
\caption{
The escape rate ($k_{\rm e}$: see text for details) as a function of cluster age 
(scaled by the half-mass relaxation time) for the models in this work 
as well as models K100-00a and K100-00b of Hurley (2007). 
\label{f:fig8}}
\end{figure}

In Figure~\ref{f:fig7} we look at how the bound cluster mass changes with cluster age 
for the various models in this work. 
We also include the K100-00a and K100-00b models of Hurley (2007) for comparison. 
As we would expect the clusters in the weaker NGC$\,6822$-like tidal field lose stars/mass 
at a slower rate compared to clusters orbiting within a galaxy that is ten times more massive. 
Also, for models M1 and M2 (where M2 is the lower of the dashed lines) we once again 
see that evolution is more rapid for the $W_0 = 7$ King model compared to the Plummer model, 
although the slopes are similar from about $5\,$Gyr onwards. 
For the NGC$\,6822$ model clusters we see that more mass is lost by any particular time if 
the cluster was initially closer to filling its tidal radius 
(noting that N1 is the upper of the solid lines). 
This corresponds to the finding of Gieles \& Baumgardt (2008) that clusters that are initially 
well within the tidal radius take a much greater number of half-mass relaxation times to lose 
half of their mass than clusters that fill their tidal radius. 
Following Giersz \& Heggie (1997) an expression for the mass-loss rate can be written as: 
\begin{equation} 
\label{e:dmdt} 
\frac{{\rm d}M}{{\rm d}t} \simeq - k_{\rm e} \frac{M}{t_{\rm rh} \ln \Lambda} 
\end{equation} 
where $\ln \Lambda$ is the Coulomb logarithm and we take $\Lambda = 0.4 N$. 
For their $N$-body models of $N = 500$ stars Giersz \& Heggie (1997) found a value 
of 1.3 for $k_{\rm e}$. 
In Figure~\ref{f:fig8} we plot $k_{\rm e}$ as a function of the number of half-mass 
relaxation times that have elapsed. 
We see that the behaviour for all of the clusters that initially  fill their tidal radii is 
similar, in particular N3 follows M1, and that $k_{\rm e}$ decreases as the clusters 
become more evolved. 
We also see that $k_{\rm e}$ is lower for clusters with smaller initial $r_{\rm h} / r_{\rm t}$, 
a result that is predicted analytically by Gieles \& Baumgardt (2008) 
and previously demonstrated by the Monte Carlo models of Spitzer \& Chevalier (1973). 
In this work we use Eq.~\ref{e:dmdt} and Figure~\ref{f:fig8} simply as a way of comparing the 
mass-loss evolution of the various models. 
A detailed examination of how dissolution times for star clusters in tidal fields can be 
expected to scale with cluster parameters can be found in Baumgardt \& Makino (2003) 
and Gieles \& Baumgardt (2008). 

We also remind the reader at this stage that the values of $r_{\rm h}$ are based on 
three-dimensional data. 
If instead we calculated the half-mass radii from a two-dimensional projection a reduction of 
about 25 per cent would result (Fleck et al. 2006). 
Furthermore, Hurley (2007) showed that projected half-light radii 
could be as much as half that of the corresponding half-mass radii 
for dynamically evolved clusters. 
In Table~\ref{t:table2} we show both the half-light (two dimensional) and 
half-mass (three-dimensional) radii for the models at an age of $12\,$Gyr. 
The errors in both values are also shown. 
Most obvious is that the errors in $r_{\rm h,l}$ are greater than in $r_{\rm h}$ 
reflecting that the former quantity fluctuates on a short timescale 
resulting from a sensitivity to the movement of massive stars in combination 
with events such as the evolution of bright giants to become dim white dwarfs. 
%{\bf [Need to make that explanation clearer!]} 
We can find particular times when $r_{\rm h,l}$ is as low as 50\% that of $r_{\rm h}$.  
However, on average the reduction is in the 30-35\% range. 
It might also be expected that the $r_{\rm h,l}$/$r_{\rm h}$ ratio would be smaller for 
more evolved clusters with a greater degree of mass-segregation (say M1 compared to N3) 
but overall we see only a weak trend for this. 

\section{Discussion}
\label{s:discus}

We have shown with our model N3 that clusters in the outer regions of a galaxy 
such as NGC$\,6822$ can naturally evolve to have large half-mass radii 
provided they begin by filling their tidal radii. 
The half-light radii in projection can exceed $15\,$pc which is within 
the range observed for extended clusters. 
Furthermore, extended clusters are observed out to $\sim 11\,$kpc in projection 
from the centre of NGC$\,6822$ which corresponds to a three-dimensional 
$R_{\rm gc}$ of $\sim 15\,$kpc. 
Our models were evolved at $R_{\rm gc} = 10\,$kpc so following 
equation~\ref{e:rtide} we can scale the tidal radius by a factor of 1.5 and 
potentially the half-mass and half-light radii by a similar factor. 
Thus half-light radii in excess of $20\,$pc are possible. 
Also consider that our models started with $N = 100\,000$ and have a mass 
of about $30\,000 \, M_\odot$ remaining at $12\,$Gyr. 
If some of the actual clusters are in fact more massive, say twice as massive, 
then following equation~\ref{e:rtide} again gives another potential 25\% increase 
in the quoted half-light radii. 
Therefore, for a dwarf irregular galaxy such as NGC$\,6822$ extended clusters 
with projected half-light radii in the range $10 - 30\,$pc can be expected to 
naturally occur (for a range of masses and orbits). 
Such clusters will evolve slowly. 
They will be dynamically young, having evolved for approximately two half-mass 
relaxation times, and will not show any signs of core-collapse. 

Any clusters in the outer regions of NGC$\,6822$ with half-light radii less than $10\,$pc 
will need to have started life relatively compact. 
Certainly they would not be expected to have filled their tidal radii initially. 
For a cluster in NGC$\,6822$ to exhibit core-collapse within a Hubble time an 
initial filling factor of less than 30\% is required, or an orbit at $R_{\rm gc}$ much less than $10\,$kpc. 
We have shown that for a fixed $R_{\rm gc}$ differences in initial sizes can translate 
to differences in half-light radii at $12\,$Gyr by the same factor. 
We have also shown that primordial binaries do not affect the evolution of the half-mass 
and half-light radii. 
This is not surprising for dynamically young clusters and has also been demonstrated 
for more evolved clusters in stronger tidal fields (Hurley 2007). 
The internal evolution characteristic that has the most potential to impact the half-light 
radius evolution is the formation of a central population of BH-BH binaries. 
Expansion of the half-light radius by this mechanism was shown by Mackey et al. (2008) 
to start taking effect after one half-mass relaxation time, with the effect increasing to 
a maximum expansion factor of two by core-collapse (comparing clusters with and without 
BH-BH binaries). 
So this could be a mild factor for the most extended clusters in NGC$\,6822$ and 
potentially play a larger role in determining the observed half-light radii of initially 
more compact clusters. 
However, this relies on a source of BHs being retained in the clusters. 
Overall we find that differences in initial cluster size are the most likely explanation 
for any measured differences in half-light radii for clusters that otherwise appear similar. 
The stochastic effect of the presence, or not, of a central population of BH-BH binaries 
can then possibly play an added role. 

Our question of ``how do similar clusters obtain different structure?'' now seemingly becomes 
``can clusters be expected to be born with different sizes?''. 
Baumgardt et al. (2010) asked a similar question after showing that Milky Way GCs 
beyond $R_{\rm gc}$ of $8\,$kpc fall into distinct groups of compact, tidally-underfilling clusters 
and more extended, tidally-filling clusters. 
They concluded it was more likely that the second group formed with large half-mass 
radii rather than evolving from a more compact state. 
Examining the size distribution of globular clusters in a variety of Local Group 
dwarf galaxies, Da Costa et al. (2009) suggest that there may be two distinct modes 
of cluster formation with typical half-light radii of $3$ and $10\,$pc, analogous 
to the findings of Baumgardt et al. (2010). 
It has been suggested by Ballesteros-Paredes et al. (2009) that tidal forces may play 
a role in determining the rate at which stars form within molecular clouds, through 
compression and possible disruption of the clouds. 
Their calculations show that the resultant star formation efficiency of a particular 
cloud depends on its position and orientation in the host galaxy. 
This model is put forward as an alternative to the theory that magnetic field strength 
and radiative feedback regulate the degree of star formation within a cloud 
(Price \& Bate 2009). 
Either way, if differences in star formation efficiency can exist between 
proto-clusters forming in the cores of molecular clouds then Baumgardt \& Kroupa (2007) 
have shown that different $r_{\rm h} / r_{\rm t}$ values can result for the bound clusters 
remaining after residual gas is expelled. 
Furthermore, simulations conducted by Elmegreen (2008) have demonstrated that 
diffuse star clusters can form directly in regions where the background tidal forces 
are low. 
Here star formation efficiency is discussed in terms of turbulence (or Mach number) 
and cloud pressure (or column density). 
Different combinations produce clusters of different densities with high-pressure 
regions making the densest clusters while clouds with low density and high Mach 
numbers give rise to diffuse clusters. 
The latter are more likely to form in the outer regions of a galaxy where low-density clouds 
are expected to reside and the diffuse clusters will not be quickly disrupted 
(Elmegreen 2008). 
So there are possibilities for variations in environment and/or the properties of molecular clouds 
producing different star formation efficiencies and a range of $r_{\rm h} / r_{\rm t}$ 
values for young star clusters. 
What is not immediately clear is whether or not clusters forming in the same 
environment can have a range of $r_{\rm h} / r_{\rm t}$ values owing to variations 
in cloud properties (size, shape, orientation) alone. 

We now turn our focus to the M31 extended cluster system. 
Our model M1 which started tidally-filling demonstrates the maximum size that 
a cluster of $M \sim 15\,000 \, M_\odot$ orbiting at $R_{\rm gc} = 10\,$kpc 
can be expected to have at an age of $12\,$Gyr. 
This is not within the extended cluster size range. 
Of course we are not claiming that our model M1 is a direct representation of a 
typical star cluster in M31. 
For a start, the mass of the model is likely to be an underestimate compared 
to actual M31 clusters. 
The extended clusters observed in M31 have masses that range from $\sim 10^4$ 
to $\sim 10^5 \, M_\odot$ (Huxor et al. 2010a: 
note that masses for ECs in NGC$\,6822$ are not yet known but are expected to 
cover a similar range). 
Thus we have direct models at the lower end of this range but could be as much 
as a factor of ten too small at the upper end. 
A factor of ten increase in mass will lead to a larger tidal radius, 
but only by a factor of two and if the internal radii scale in the same way this 
gives a projected half-light radius in the range of $10\,$pc (see Table~\ref{t:table2}). 
This is right at the low-end of the extended cluster size range. 
However, this still does not help when we are looking at two clusters of the same mass 
(or luminosity) at the same  $R_{\rm gc}$ and with very different half-mass radii, 
as is observed in the inner regions of M31 (Huxor et al. 2010a). 
Thus it is hard to believe that extended clusters observed in the inner regions of M31, 
which have half-light radii of about $30\,$pc, evolved in-situ. 
It is possible that extended clusters observed in the inner regions are actually further 
out and only appear in the centre owing to projection effects. 
Such an effect is not noticeable from the horizontal-branch levels of the few extended 
clusters that have good colour-magnitude diagrams (Mackey et al. 2006) and if there 
is a projection effect pushing these clusters into the foreground there is no reason 
why it shouldn't similarly apply to the compact clusters as well. 
On the other hand we expect that compact clusters will be more likely to survive in the 
inner parts of galaxies than extended clusters and radial distributions of GCs in 
galaxies show that numbers increase towards the centre (see Brodie \& Strader 2006, 
for example). 
So for compact clusters at least it is reasonable to believe that there is a sizeable 
population residing in the inner regions of galaxies such as M31, regardless of 
projection effects. 

It is also true that for our models we have used an  $R_{\rm gc}$ that for M31 
represents only the very inner edge of the extended cluster locations. 
As discussed above, and shown by equation~\ref{e:rtide}, in a smooth tidal field 
the tidal radius of a cluster scales linearly with $R_{\rm gc}$. 
So the maximum possible $r_{\rm h}$ will also increase for clusters observed 
further out than $10\,$kpc. 
The M31 extended clusters are observed out to $R_{\rm gc}$ of $100\,$kpc 
and beyond which means a potential factor of ten increase in size for clusters 
in the outer regions. 
We note though that other factors such as the relaxation timescale will be 
affected by a change in cluster size so it is not necessarily a simple case 
of scaling up a model value at a particular age. 
To gain some insight we can turn to 
Mackey \& van den Bergh (2005) who have indeed shown a clear trend of increasing 
$r_{\rm h}$ with increasing $R_{\rm gc}$ for Milky Way globular clusters 
(see their figure~9). 
However, the relation is not linear: 
going from $R_{\rm gc}$ of 10 to $100\,$kpc gives a factor of $2 - 3$ 
increase in the typical cluster size (as measured by $r_{\rm h}$). 
If we assume a similar scaling for M31 then we can expect to find clusters 
with  $r_{\rm h,l} \sim 30\,$pc in the outer regions. 
It is particularly interesting to note that the $r_{\rm h,l}$ values derived for the M31 
extended clusters (Huxor et al. 2010a) show no obvious correlation with $R_{\rm gc}$. 
In combination with our model results 
this infers that the inner M31 extended clusters either resided at more distant locations 
for a large part of their lives before moving to their current locations or they 
were likely accreted as part of one or more dwarf galaxies 
(perhaps similar in nature to NGC$\,6822$ or the dwarf elliptical Scl-dE1). 
This latter scenario fits with the picture sketched by Da Costa et al. (2009) where 
the formation of extended clusters can be expected to be more common in dwarf galaxies 
where tidal effects are less disruptive and that extended clusters seen in larger 
galaxies, M31 for example, were accreted in dwarfs that subsequently disrupted. 
Adding to this argument is evidence of a genuine physical association between the 
extended clusters (as well as most compact clusters) and coherent stellar streams in 
the halo of M31 (Mackey et al. 2010). 

Our models so far have been evolved within a smooth background 
potential represented by a point-mass galaxy.
This simplification may cause discrepancies when comparing our models 
to clusters that evolve in host galaxies with distinct structures. 
In particular, both the Milky Way and M31 exhibit pronounced 
bulge, disk and halo components which means that the effects 
of bulge and disk shocking need to be considered. 
Gnedin, Lee \& Ostriker (1999) have demonstrated with Fokker-Planck simulations 
that gravitational shocks caused by the time-varying tidal forces as clusters pass 
near the central Galactic bulge or through the Galactic disk can significantly 
reduce core-collapse and dissolution times. 
The cluster binding energy is reduced by the shocks so that diffuse clusters are 
the most susceptible to disruption. 
However, the effect of the shocks is strongly dependent on the Galactocentric radius. 
Bulge shocking dominates cluster evolution for $R_{\rm gc}$ of $2\,$kpc or less 
(Gnedin \& Ostriker 1997) but drops off quickly away from the central nucleus, 
while disk shocking has a minimal effect for $R_{\rm gc}$ beyond about $8\,$kpc 
(Vesperini \& Heggie 1997) owing to lower disk density and less frequent passages. 
%Also, Pichardo, Martos \& Moreno (2004) showed that the presence of a central Galactic 
%bar only affects clusters within $3\,$kpc of the Galactic centre. 
To check this we have repeated model M1 using a three-dimensional Galactic potential: 
a point-mass bulge, a Miyamoto \& Nagai (1975) disk (with $a = 4\,$kpc and $b = 0.5\,$kpc) 
and a logarithmic halo potential 
(see Aarseth 2003 and Praagman, Hurley \& Power 2010 for the {\tt NBODY6} implementation). 
Masses of $5 \times 10^{10}$ and $1.5 \times 10^{10} \, M_\odot$ were chosen for the bulge 
and disk, respectively (Xue et al. 2008), with the halo mass chosen so that the mass 
interior to $10\,$kpc was the same as for model M1. 
The cluster was evolved at $R_{\rm gc} = 10\,$kpc perpendicular to the disk and 
there were no obvious changes in evolution of cluster mass or the structural parameters 
compared to M1. 
Thus we are confident that our use of a point-mass galaxy has little bearing on our 
results for clusters evolving at $R_{\rm gc}$ of $10\,$kpc or greater. 
Conversely, as suggested by Da Costa et al. (2009), the action of tidal shocks makes 
it unlikely that extended clusters would be found at smaller galactic radii in large disk 
galaxies such as M31. 
A full investigation of the effect of three-dimensional galaxy potentials on the 
evolution of star clusters will be the focus of future work, noting that the scale-length 
of the disk of M31 ($\sim 6\,$kpc) is larger than for the Milky Way disk 
($\sim 4\,$kpc: Yin et al. 2009). 

Another avenue for future study is the effect of the initial cluster density profile 
on the subsequent evolution. 
Trenti, Vesperini \& Pasquato (2010) find that King models with different 
initial concentrations evolve to a similar structural state within a strong tidal field. 
However, the differences we observed between our models M1 and M2 
(evolved in the M31-like tidal field) means that the effect of changing the initial core-density 
should be investigated for models evolving in a weaker NGC$\,6822$-like tidal field. 

We should also note that the results and conclusions we draw regarding extended clusters 
are based on the assumption that these are essentially normal star clusters where the 
dynamical processes are not influenced by the presence of any significant dark matter component. 
If extended clusters are found to contain dark matter then our results to date are not applicable. 
The only direct constraints on this possibility thus far are the measurements for one M31 extended 
cluster reported by Collins et al. (2009). 
Their derived mass-to-light ratio of $M / L = 6.7^{+15}_{-6.7} \, M_\odot / L_\odot$ rules out 
very large ratios as seen in most dwarf spheroidals. 
As stated by Collins et al. (2009) the value is most consistent with a typical star cluster. 
However, the possibility of $M/L$ as high as 20 still exists so the issue of whether or not extended 
clusters contain dark matter is yet to be resolved. 
The case against dark matter is strengthened by 
Jordi et al. (2009) who measured radial velocities of red giants in Palomar 14, 
a diffuse GC in the outer regions of the Milky Way, to derive a modest ratio 
of $M / L \sim 2$. 
This is compatible with the value expected for a normal stellar initial mass function 
with no dark matter.

%Note that the locations of the observed ECs are projected distances so that the actual 
%3-d $R_{\rm gc}$ will be greater. 
%So $R_{\rm gc} \simeq 15\,$kpc would have been a better choice to more closely correspond to the 
%outer EC in NGC$\,6822$ and the most inner ECs in M31
%(wish I had gone with this originally, however at least $10\,$kpc may be a good representation 
%of the EC we observed with Keck which is at a projected $6\,$kpc). 
%This means that $r_{\rm h}$ and $r_{\rm h,l}$ results can be scaled by $\sim 1.5$ if desired 
%and thus it is possible to get $r_{\rm h,l} > 20\,$pc for the NGC$\,6822$ model clusters, 
%corresponding to the outermost observed. 

%The survivability of ECs. 
%Discuss Figure~\ref{f:fig5} in terms of the ``survival triangle'' of Gnedin \& Ostriker (1997) and 
%the recent paper by Baumgardt et al. (2009), as well as the $M$--$r_{\rm h}$ observations plot ...
%Should note here that different combinations of initial $M$ and $r_{\rm h}$ can lead 
%to the same final $M$ at a later time. 

%Discuss how our models explore and extend ideas from 

Finally, our models build on a number of important results from previous work 
regarding general star cluster evolution. 
A prime example is the work of 
K\"{u}pper, Kroupa \& Baumgardt (2008) who used $N$-body models to demonstrate 
that post-collapse clusters 
asymptotically evolve on to a common sequence where the $r_{\rm h} / r_{\rm t}$ ratio 
depends only on the mass remaining in the cluster and not on the initial conditions. 
They considered open clusters starting with up to $N \simeq 32\,000$ stars and different binary 
fractions, density distributions, $r_{\rm h}$ and  $R_{\rm gc}$ values. 
Our models extend this work to include a distribution of stellar masses and 
stellar evolution for $N = 100\,000$. 
In doing so we confirm the findings of K\"{u}pper, Kroupa \& Baumgardt (2008). 
The common sequence shows a $r_{\rm h} / r_{\rm t}$ ratio that starts in the 
range of $0.15 - 0.2$ and increases slightly with decreasing mass. 
Related to this the same authors found that clusters in a tidal field show an 
equilibrium half-mass radius after core-collapse. 
This was $2\,$pc for their standard set of models, noting that the particular value 
will increase for higher initial mass and larger $R_{\rm gc}$. 
Our models M1 and M2 have a similar tidal tidal field but larger initial mass. 
These models also show an equilibrium half-mass radius and as expected it is 
larger, $\sim 5\,$pc, with the inclusion of stellar evolution likely contributing 
to some of the increase. 
This corresponds to a projected half-light radius of about $3\,$pc which fits with 
the typical value observed for compact GCs in the Milky Way (Baumgardt et al. 2010) 
and M31, and indeed almost all galaxies where we see GCs 
(Da Costa et al. 2009, for example). 
Of our models in the weaker tidal field only N1 gets close to core-collapse 
and indeed has a larger half-mass radius at this point, $\sim 10\,$pc, but in this 
case appears to still be rising.

\section{Conclusions}
\label{s:summ}

Using a direct $N$-body code we have followed the evolution of 
star clusters with different initial sizes in a tidal field appropriate 
for a galaxy such as the dwarf irregular NGC$\,6822$ or the 
Large Magellanic Cloud. 
We also looked at the effect of increasing the galaxy mass by a 
factor of ten, appropriate for larger galaxies such as M31 and the Milky Way, 
on the evolution of clusters which initially fill their tidal radii. 
Our main findings can be summarised as: 
\begin{itemize}
\item extended clusters with projected half-light radii of up to $30\,$pc 
          in galaxies such as NGC$\,6822$ 
          can result from standard star cluster evolution provided that the 
          clusters are close to filling their tidal radii initially; 
\item internal evolution factors such as the proportion of primordial binaries or 
          the formation of BH-BH binaries do not 
          produce enough expansion so formation as an extended object followed 
          by standard cluster evolution is the most likely path for the extended 
          clusters recently observed; 
\item observations of clusters of similar mass and age but significantly 
          different half-light radii most likely indicate that the clusters started 
          life filling their tidal radii by different factors; 
\item in a galaxy with a strong tidal field, such as M31, the maximum 
          half-light radius expected for clusters that evolve at a galactocentric 
          radius of $10\,$kpc is of the order of $10\,$pc; 
\item extended clusters could form naturally in the outer regions of M31 
          (at $\sim 100\,$kpc) but any extended clusters observed in the inner 
          regions likely formed in one or more dwarf galaxies that were 
          subsequently accreted by M31.  
\end{itemize}

\section*{Acknowledgments}
We are indebted to Sverre Aarseth and Keigo Nitadori for creating and 
maintaining the GPU library for {\tt NBODY6}. 
JRH also wishes to thank Ben Barsdell and David Barnes for assistance 
with GPU usage at Swinburne, as well as Annie Hughes for informative 
discussions regarding molecular clouds. 
We also thank the referee for a number of insightful comments.

\label{lastpage}

\end{document}